\documentclass[%
  showkeys,amsmath,amssymb,
aps,
pra,
]{revtex4-2}
\usepackage{booktabs}
\usepackage{graphicx}
\usepackage{dcolumn}
\usepackage{bm}
\usepackage{hyperref}

\renewcommand{\doi}[1]{\href{https://doi.org/#1}{DOI: \nolinkurl{#1}}}

\graphicspath{ {./figs/} }
\usepackage{orcidlink}
\begin{document}
\title{What visibility-graph curvature measures in Forbush decreases: a calibration on the two largest events of solar cycle 25}

\author{D. Sierra-Porta\orcidlink{0000-0003-3461-1347}}
\email{dporta@utb.edu.co}
\affiliation{Universidad Tecnológica de Bolívar. Grupo de Investigación Gravitación y Matemática Aplicada - GIGMA \& Grupo de Investigación Física Aplicada y Procesamiento de Imágenes y Señales - FAPIS., Parque Industrial y Tecnológico Carlos Vélez Pombo Km 1 Vía Turbaco, Cartagena de Indias, 130010, Bolívar, Colombia}

\begin{abstract}
Discrete curvatures of visibility graphs are increasingly used as
descriptors of geophysical time series, but their response to a transient
has rarely been calibrated against the simplest properties of that
transient. We carry out this calibration for the nodal Forman--Ricci
curvature $\mathcal{F}$ of natural visibility graphs applied to
neutron-monitor records of the Gannon storm (2024 May 10--11) and the
2025 June 1 Forbush decrease (FD), the two largest FDs of solar cycle 25,
using 30-min NMDB data from six NM64 stations ($R_c \simeq 0$--$7$~GV),
i.e.\ twelve station--event series. Curvature collapses at FD onset in
every series (Cliff's $\delta = 0.63$--$1.0$), but a 6-h running standard
deviation separates onset from quiet time more strongly (mean $\delta =
0.96$ vs $0.89$; mean block-permutation displacement $z = -27$ vs $-14$).
Injecting synthetic decreases into the real pre-event records of each
station shows that the curvature extreme follows $|\mathcal{F}_{\min}|
\propto (A/\sigma)^{0.50}$ ($R^2 = 0.88$), where $A$ is the FD amplitude
and $\sigma$ the station's quiet-time noise, and is independent of the
duration of the main phase (exponent $0.006$ over 3--36~h). With $A$,
$\sigma$ and the main-phase duration of each series as the only inputs,
the calibration predicts the twelve observed extremes with standardized
residuals of mean $-0.26$ and standard deviation $0.94$, and the observed
extremes correlate with $A/\sigma$ at Spearman $\rho = 0.95$. Within the
precision allowed by these data, nodal Forman--Ricci curvature of the
whole-series visibility graph carries no information about these FDs
beyond their amplitude relative to station noise; in particular, it does
not distinguish the single-step Gannon decrease from the two-step June
2025 decrease. We argue that any morphological or physical
interpretation of visibility-graph curvature in cosmic-ray data should
first be tested against this baseline.
\end{abstract}


\maketitle

\section{Introduction}

Visibility graphs map a time series onto a network by linking samples that
``see'' each other over the intervening record \citep{lacasa2008,luque2009},
and the resulting graphs inherit properties of the series such as its
roughness \citep{lacasa2009}. Discrete Ricci curvatures defined on these
graphs \citep{forman2003,sreejith2016,samal2018} have been proposed as
markers of regime changes in complex systems \citep{sandhu2016}.
Visibility graphs have been applied to hurricane records, where the hubs
of the graph coincide with the most active seasons \citep{elsner2009},
and, in heliophysics, to solar-wind velocity \citep{suyal2014}, to sunspot
and solar magnetic-field records spanning several cycles
\citep{zou2014,zuritavalencia2023}, and to magnetic fluctuations in
simulated non-thermal plasmas \citep{acostatripailao2021,saldivia2024};
\citet{zou2019} review these methods. Applications to neutron-monitor
records are fewer. Visibility-graph and curvature descriptors of daily and
monthly records have been related to the geomagnetic cutoff rigidity
\citep{sierraporta2024,sierraporta2025}, and visibility-graph curvature has
been proposed as a reader of the May 2024 Forbush decrease
\citep{sierraporta2026}. The usual argument is that a transient reshapes the local connectivity of
the graph and therefore its curvature.

That argument is correct, but it leaves open what the curvature adds. A
Forbush decrease is a large, fast drop in the count rate followed by a slow
recovery \citep{lockwood1971,cane2000}. Any statistic sensitive to
amplitude or local variance will respond to it. \citet{donner2012} warned
that network measures of geophysical series can reflect elementary
properties such as series length, sampling and non-stationarity rather
than the underlying dynamics. The question that matters
for using curvature as a descriptor is whether its response contains
information beyond the size of the drop relative to the noise of the
detector, for instance about the duration of the main phase or the
distinction between single-step (sheath-dominated) and two-step
(sheath plus ejecta) decreases \citep{cane2000,jordan2011}. If it does
not, curvature values cannot be read as a morphological fingerprint, and
differences between stations or events reflect amplitude and detector
noise.

We address this question with a calibration rather than a new detection
exercise. The test bed is the pair of largest FDs of solar cycle 25,
recorded by the same six neutron monitors: the Gannon storm of 2024 May,
whose main phase lasted about 6.5~h \citep{hayakawa2025}, and the 2025 June
event, a compound decrease driven by interacting CMEs that developed in two
steps over about 20~h \citep{sevan2025}. Their different tempos make them a
natural check on whether curvature resolves morphology. We first compare
the onset response of curvature with that of elementary statistics on the
same data, then derive the response of curvature to synthetic decreases
injected into the real quiet records of each station, and finally ask
whether the observed series depart from that calibration.

\section{Data and graph construction}

\begin{figure}
\centering
\includegraphics[width=\textwidth]{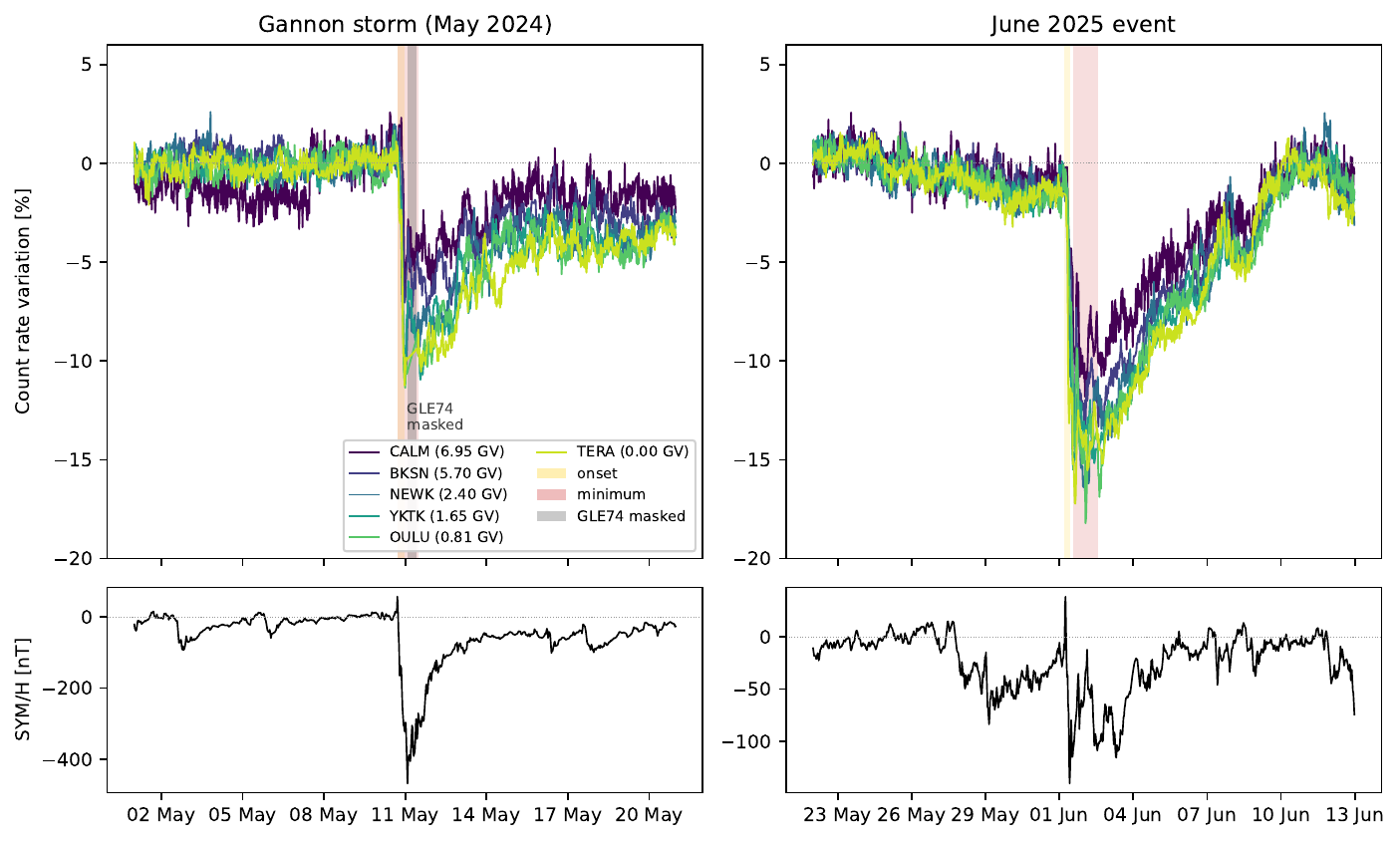}
\caption{Normalized count-rate profiles of the six stations (top) and OMNI
SYM/H (bottom) for the Gannon storm (left) and the June 2025 event (right).
Shading marks the onset and minimum intervals and the masked GLE74
interval.}
\label{fig:profiles}
\end{figure}

We use pressure- and efficiency-corrected 30-min count rates from the
Neutron Monitor Database \citep[NMDB;][]{mavromichalaki2011} for 2024 May
1--20 and 2025 May 22--June 12. After gap and spike screening, six NM64
stations present and clean in both windows were retained
(Table~\ref{tab:stations}); isolated outliers ($>5$ robust-$\sigma$ from a
6-h running median, $\sim$0.5\% of points) were interpolated. Series are
expressed as percentage deviations from pre-event baselines (2024 May 7
00:00--May 10 12:00 and 2025 May 22 00:00--May 27 12:00 UT). The GLE74
interval (2024 May 11 02:00--10:00 UT) \citep{papaioannou2025} is masked
and linearly interpolated at OULU and TERA. Shock arrivals are taken at
2024 May 10 17:00 UT \citep{hayakawa2025} and 2025 June 1 05:00 UT
\citep{sevan2025}. We define the \emph{pre} window as the start of the
record up to the shock and the \emph{onset} window as the 6~h following
the shock. Figure~\ref{fig:profiles} shows the profiles. The amplitude $A$
of each series is its maximum depression; it ranges from 6.6\% (CALM,
Gannon) to 18.2\% (OULU, June 2025).

\begin{table}
\centering
\caption{Neutron monitors used, and the quiet-time noise level $\sigma$
of each (see Sect.~\ref{sec:calib}).}
\label{tab:stations}
\begin{tabular}{lcccc}
\toprule
Station & $R_c$ [GV] & Altitude [m] & Detector & $\sigma$ [\%] \\
\midrule
CALM & 6.95 & 708  & 15-NM64 & 0.66 \\
BKSN & 5.70 & 1700 & 6-NM64  & 0.34 \\
NEWK & 2.40 & 50   & 9-NM64  & 0.38 \\
YKTK & 1.65 & 105  & 24-NM64 & 0.26 \\
OULU & 0.81 & 15   & 9-NM64  & 0.38 \\
TERA & 0.00 & 32   & 9-NM64  & 0.35 \\
\bottomrule
\end{tabular}
\end{table}

\begin{figure}
\centering
\includegraphics[width=\textwidth]{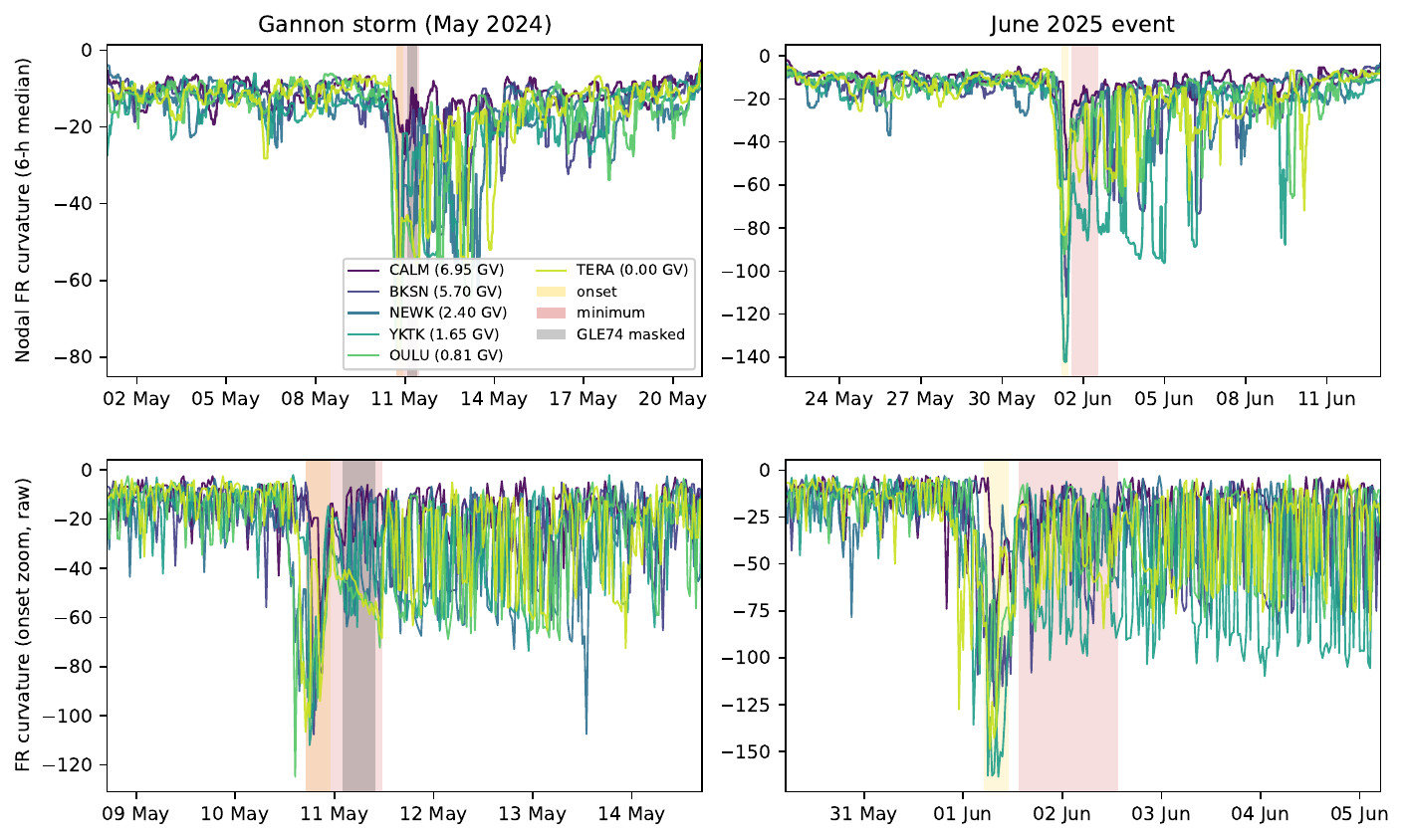}
\caption{Nodal Forman--Ricci curvature of the twelve series: 6-h running
median over the full window (top) and raw values around onset
(bottom).}
\label{fig:curv}
\end{figure}

For each of the twelve station--event series we build the natural
visibility graph (NVG) of the full record ($n = 960$ and $1056$ samples):
samples $i<j$ are linked if
\begin{equation}
y_k < y_i + (y_j - y_i)\,\frac{k-i}{j-i} \qquad \forall\; i<k<j
\label{eq:nvg}
\end{equation}
\citep{lacasa2008}. The combinatorial Forman--Ricci curvature of an edge is
$F(u,v) = 4 - k_u - k_v$ \citep{forman2003,sreejith2016}, with $k_u$ the
degree, and we assign to each sample the nodal curvature
\begin{equation}
\mathcal{F}(u) = \frac{1}{k_u} \sum_{v \in \mathcal{N}(u)} F(u,v)
 = 4 - k_u - \langle k \rangle_{\mathcal{N}(u)} .
\label{eq:nodal}
\end{equation}
The second form shows what $\mathcal{F}$ responds to. A deep, fast drop
produces samples that see across large parts of the record and become
hubs; their neighbors inherit a large $\langle k \rangle_{\mathcal{N}}$
and therefore a strongly negative $\mathcal{F}$. Figure~\ref{fig:curv}
shows the resulting curvature series. We summarize each series by two
numbers: the extreme $|\mathcal{F}_{\min}|$ over the main and minimum
phases, and the mean $|\mathcal{F}|$ over the onset window.

\section{Methods}

\subsection{Comparison with elementary statistics}

To measure how well a statistic separates onset from quiet time, we use
Cliff's $\delta$ between its pre and onset values, and a placement test in
which the mean over the onset window (13 samples) is compared with the
means of all contiguous 13-sample windows of the pre interval
($\sim$450 per series). The placement test preserves the autocorrelation
of each statistic exactly; we report the displacement $z$ of the onset
mean in units of the standard deviation of the null means. Besides
$\mathcal{F}$, we evaluate the negative degree $-k$, a 6-h centered running
standard deviation and running range of the count rate, the absolute
first difference $|\Delta y|$, and the count rate itself. Statistics are
signed so that positive $\delta$ and negative $z$ indicate the expected
onset response.

\subsection{Synthetic calibration}
\label{sec:calib}

For each station we build a quiet record of $\sim$950 samples by
concatenating the two pre-event intervals, each with its mean removed,
and apply random circular shifts to generate four realizations. The noise
level is quantified as $\sigma = \mathrm{std}(\Delta y)/\sqrt{2}$, the
standard deviation of the white component of the quiet record
(Table~\ref{tab:stations}); it differs by a factor of 2.5 among the six
stations, reflecting counting statistics and detector size. Into each
realization we inject a decrease that falls linearly to $-A$ over a main
phase of duration $T_{\rm main}$, starting at a random position, and
recovers exponentially with an e-folding time of 3 days, for $A \in
\{3, 5, \dots, 21\}\%$ (nine values) and $T_{\rm main} \in \{3, 6, 12,
24, 36\}$~h. This gives 1080 synthetic series. The design is a surrogate-data test
\citep{theiler1992,schreiber2000} with a physically specified null
hypothesis: a decrease with no internal structure, of given amplitude and
duration, superposed on the station's own quiet-time fluctuations. For each we compute
$|\mathcal{F}_{\min}|$ over the main phase plus the following 12~h, the
mean $|\mathcal{F}|$ over the 6~h after the start of the injection, and
Cliff's $\delta$ between the pre-injection and onset values, exactly as
for the observed series. We then fit $\log|\mathcal{F}_{\min}|$ and the
log of the onset mean on $\log(A/\sigma)$ and $\log T_{\rm main}$.

To compare with the observations, we fit the same regression separately
for each station (on $\log A$ and $\log T_{\rm main}$, so that the
station's own noise is built in) and predict the twelve observed values
from the observed $A$ and $T_{\rm main}$ (time from shock to the
station's own minimum). The standardized residual $z_{\rm res} =
[\log(\mathrm{obs}) - \log(\mathrm{pred})]/s$, with $s$ the residual
standard deviation of the station's synthetic fit, measures whether an
observed series departs from what a structureless ramp with the same
amplitude and noise would produce.

\section{Results}

\subsection{Onset response compared with elementary statistics}

Table~\ref{tab:stats} summarizes the twelve series. Curvature separates
onset from quiet time in every case, with Cliff's $\delta$ between 0.63
and 1.0, but it is not the strongest separator. The 6-h running standard
deviation and running range reach mean $\delta$ of 0.96 and 0.95, and
their minima over the twelve series (0.73 and 0.71) exceed that of
curvature. In the placement test, which accounts for autocorrelation, the
running standard deviation displaces the onset mean by $z = -27$ on
average against $-14$ for curvature, and exceeds curvature in $|z|$ in ten
of twelve series. The negative degree alone has a lower $\delta$ than
curvature (0.54) but a larger mean $|z|$ (22 against 14), because its
quiet-time variance is smaller. The running statistics are centered and
thus use 3~h of future data; this is the fair comparison, since the
whole-series NVG uses the entire record, but a strictly causal 6-h
running standard deviation still reaches mean $\delta = 0.73$.

\begin{table}
\centering
\caption{Onset versus pre-event separation for six statistics over the
twelve station--event series: Cliff's $\delta$ (mean and minimum) and
placement-test displacement $z$ (mean and least negative).}
\label{tab:stats}
\begin{tabular}{lcccc}
\toprule
Statistic & $\langle\delta\rangle$ & $\delta_{\min}$ & $\langle z\rangle$ & $z_{\max}$ \\
\midrule
6-h running standard deviation & 0.96 & 0.73 & $-27.1$ & $-7.3$ \\
6-h running range              & 0.95 & 0.71 & $-23.2$ & $-5.3$ \\
Nodal FR curvature $\mathcal{F}$ & 0.89 & 0.63 & $-14.3$ & $-5.8$ \\
Count rate                     & 0.55 & $-0.47$ & $-5.5$ & $+1.0$ \\
Negative degree $-k$           & 0.54 & 0.33 & $-22.0$ & $-9.0$ \\
$-|\Delta y|$                  & 0.53 & $-0.15$ & $-11.2$ & $-0.3$ \\
\bottomrule
\end{tabular}
\end{table}

\subsection{Calibration law}

\begin{figure}
\centering
\includegraphics[width=\textwidth]{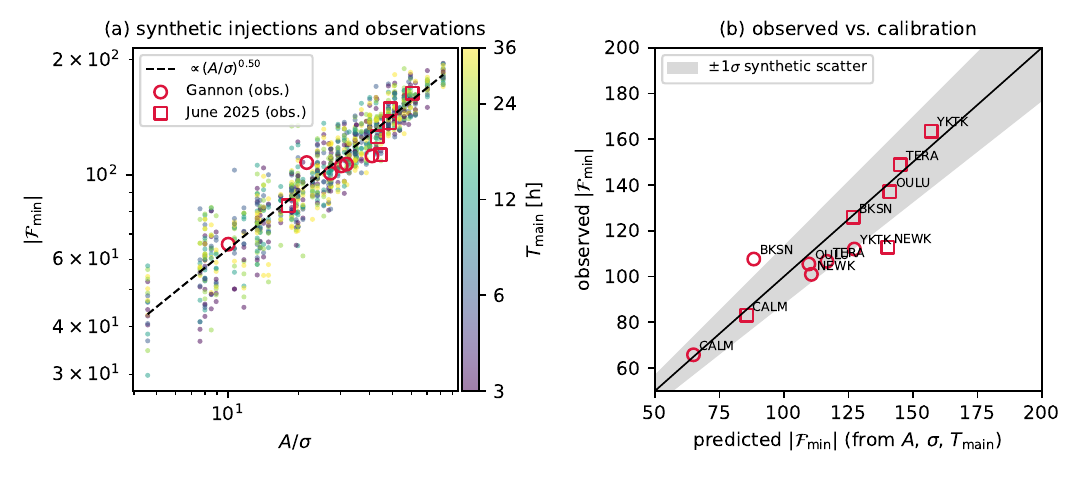}
\caption{(a) Curvature extreme of 1080 synthetic decreases injected into
the real quiet records of the six stations, against amplitude relative
to station noise, colored by main-phase duration; the dashed line is the
fit of Eq.~(\ref{eq:law}). Open symbols are the twelve observed series.
(b) Observed curvature extreme against the value predicted by each
station's calibration from its observed $A$ and $T_{\rm main}$; the band
is the $\pm1\sigma$ scatter of the synthetic fits.}
\label{fig:calib}
\end{figure}

Across the 1080 synthetic series (Fig.~\ref{fig:calib}a),
\begin{equation}
|\mathcal{F}_{\min}| \propto \left(\frac{A}{\sigma}\right)^{0.50}
T_{\rm main}^{\,0.006}, \qquad R^2 = 0.88,
\label{eq:law}
\end{equation}
with a residual scatter of 0.12 in natural log. The onset mean follows
$(A/\sigma)^{0.71}\,T_{\rm main}^{0.013}$ with $R^2 = 0.78$. The
dependence on the duration of the main phase is negligible in both cases:
changing $T_{\rm main}$ from 3 to 36~h changes $|\mathcal{F}_{\min}|$ by
less than 2\%, compared with a factor of about 4 across the range of
$A/\sigma$ explored ($A/\sigma \simeq 5$--$80$). The onset effect size also depends on $A/\sigma$
alone: the synthetic Cliff's $\delta$ rises from 0.54 at $A = 3\%$ to
$\geq 0.85$ for $A \geq 7\%$, so the large effect sizes observed in the
two events are those expected of any decrease of their size.

\subsection{Observed series against the calibration}

The observed extremes fall on the synthetic relation
(Fig.~\ref{fig:calib}). Their correlation with $A/\sigma$ is $\rho =
0.95$, against $\rho = 0.75$ with $A$ alone: most of the scatter of the
raw amplitude--curvature relation is station noise. Predicted from each
station's calibration, the twelve standardized residuals have mean $-0.26$
and standard deviation $0.94$, with the largest departure at BKSN during
Gannon ($z_{\rm res} = 1.91$) and NEWK during June 2025 ($-1.88$). For the
onset mean the residuals have mean $-0.11$ and standard deviation $0.81$
(largest $|z_{\rm res}| = 1.62$). Both distributions are what one expects
from the synthetic scatter alone. The two events are not distinguished:
the June 2025 series, whose main phase developed in two steps, are
predicted as well by a single linear ramp as the Gannon series are.

\section{Discussion}

The results identify what nodal Forman--Ricci curvature of a
whole-series NVG measures in a Forbush decrease: the depth of the drop in
units of the detector noise, compressed by a square root. The mechanism
follows from Eq.~(\ref{eq:nodal}). In the observed series the
highest-degree nodes are the samples at the upper edge of the drop, around
the shock and in the first hours of the main phase, which overlook the
whole depressed interval that follows. The curvature minimum falls on one
of the three highest-degree nodes in all twelve series, and in eleven of
them at the upper edge of the main drop; the exception, NEWK during
Gannon, is the secondary decrease of May 13. That the hubs of a natural
visibility graph mark the dominant excursions of a series is known from
other geophysical records \citep{elsner2009}. The number of depressed
samples a hub can see grows with the depth of the drop relative to the
quiet fluctuations, which would otherwise block the lines of sight, and
does not depend on how fast the drop occurs as long as it is resolved, which is why $T_{\rm main}$ drops out. The
square-root exponent is an empirical result of the injections; we have
not derived it analytically.

Three consequences follow. First, the rigidity ordering of the curvature
response across stations, and the larger response in June 2025 than in
Gannon, are inherited from the amplitudes and from the noise of each
detector. They are not independent evidence about the rigidity
dependence of the modulation. Second, curvature does not act as a
morphological fingerprint of these events: the single-step and two-step
decreases are indistinguishable once $A$ and $\sigma$ are fixed. Third, as
an onset detector it is outperformed by a running standard deviation,
which is simpler to compute and to interpret. The amplitude-blind
horizontal visibility graph, whose onset effect sizes are small for the
same series ($\delta = -0.09$ to $0.56$), points the same way: the
response lives in the amplitude geometry, which is to say in $A/\sigma$.

The calibration also bears on comparisons of mean curvature between
windows of different length. The mean nodal curvature of an NVG drifts
with graph size even for white noise, an instance of the dependence of
visibility-graph measures on series length discussed by
\citet{donner2012}. With the Forman curvature including
the triangle term, we obtain means of $-2.55$, $-2.70$ and $-2.92$ for
white-noise series of 300, 500 and 1300 samples. Differences of this
size between baseline, onset and recovery windows of unequal length,
such as those reported for the May 2024 event at Jungfraujoch
\citep{sierraporta2026}, are therefore consistent with a size effect plus
the amplitude effect described here, and do not by themselves indicate a
change of dynamical regime.

These conclusions have limits. The sample is two events at 30-min
resolution; the synthetic profile is a linear ramp with exponential
recovery and does not include the precursory pre-decrease, anisotropy or
the second step; and the noise model is the real quiet record, which
includes diurnal variation but not the enhanced turbulence of disturbed
periods. A residual that is consistent with zero in twelve series does
not exclude a small morphological signal that a catalog of hundreds of
events \citep{belov2009,richardson2010} could resolve. The calibration in Eq.~(\ref{eq:law}) provides the
baseline for that test: a physical interpretation of curvature requires a
residual that survives after $A/\sigma$ is removed. Curvature defined on
networks among stations rather than within a single series, where it
would measure the coherence of the response across the monitor network,
is a different object and is not addressed by this calibration, although
it should be held to the analogous baseline of mean inter-station
correlation.

\section{Conclusions}

For the twelve station--event series of the two largest Forbush decreases
of solar cycle 25, nodal Forman--Ricci curvature of the natural
visibility graph responds to the decrease as $(A/\sigma)^{0.5}$,
independently of main-phase duration, and the observed values are
reproduced within synthetic scatter by structureless ramps of the same
amplitude injected into the same detectors' quiet records. A 6-h running
standard deviation separates onset from quiet time more strongly. Within
the precision of these data, the curvature carries no information on
the morphology or rigidity structure of these events beyond the
amplitude relative to station noise. We suggest that applications of
visibility-graph curvature to cosmic-ray and other geophysical series
report this calibration, and the comparison with elementary statistics,
before attaching a physical meaning to curvature values.

\section*{Data and code availability}
Neutron monitor data are from NMDB (\url{https://www.nmdb.eu});
interplanetary data from the OMNI database
(\url{https://omniweb.gsfc.nasa.gov}). Analysis code and derived products
are available from the authors on request.

\section*{Acknowledgments}
We acknowledge the NMDB database (\url{www.nmdb.eu}), founded under the
European Union's FP7 programme (contract no.\ 213007). OMNI
data were obtained from the GSFC/SPDF OMNIWeb interface. We also thank the Research and Innovation Directorate of the Technological University of Bolívar for their continued support in research efforts.

\bibliographystyle{apsrev4-2}

\bibliography{main}

\end{document}